\documentclass[aps,pre,amsfonts,showpacs,unsortedaddress]{revtex4}
\usepackage[english]{babel}
\usepackage[latin1]{inputenc}
\usepackage[dvips]{graphicx}
\usepackage{amsmath}
\usepackage{amssymb}
\usepackage{epsfig}
\usepackage{array}

\newcommand{\Label}[1]{\label{#1}}   
\newcommand{\Bibitem}[1]{\bibitem{#1}}   
\newcommand{\be}{\begin{equation}}
\newcommand{\ee}{\end{equation}}
\newcommand{\ba}{\begin{eqnarray}}
\newcommand{\ea}{\end{eqnarray}}

\newcommand{\nn}{\nonumber\\}
\newcommand{\Ref}[1]{(\ref{#1})}
\newcommand{\av}[1]{\langle #1\rangle}  
\newcommand{\half}{\textstyle{\frac{1}{2}}}
\newcommand{\br}{{\bf r}}
\newcommand{\bv}{{\bf v}}
\newcommand{\bk}{{\bf k}}
\newcommand{\bV}{{\bf V}}
\newcommand{\bn}{{\bf n}}
\newcommand{\bm}{{\bf m}}
\newcommand{\bzero}{{\bf 0}}
\newcommand{\bnabla}{\mbox{\boldmath$\nabla$}}

\renewcommand{\tilde}{\widetilde}
\renewcommand{\hat}{\widehat}

\begin{document}

\title{ Universal power law  tails of time correlation functions}
\author{M. H. Ernst} \affiliation{ CNLS, Los Alamos National Laboratory, Los Alamos NM 87545 and Institute
for Theoretical Physics, University of Utrecht, Princetonplein 5,
P.O. Box 80.195, 3508 TD  Utrecht, The Netherlands}

\date{\today}

\begin{abstract}
The universal power law tails of single and multi-particle time correlation functions are derived from a unifying point of view, solely using the hydrodynamic modes of the system. The theory applies to general correlation functions, and to systems more general than classical fluids. Moreover it is argued that the collisional transfer part of the stress-stress correlation function in dense classical fluids has the same long time tail $\sim t^{-1-d/2}$ as the velocity autocorrelation function in Lorentz gases.  

\end{abstract}


\maketitle

The long time tail (LTT) of the velocity autocorrelation function 
$\av{v_{x} (t) v_x (0)} \sim t^{-d/2}$ in $d-$dimensional fluids in
thermal equilibrium is the prototypical example of time correlation functions in fluids showing universal power law behavior, independent of the
details of the inter-particle interactions \cite{LTT-sim,EHvL-JSP-MC}. The
goal of this paper is to present, from a unified point of view, a
quantitative description of the asymptotic decay of
single  and multi-particle time correlation functions (TCF's).
The starting point is the mode coupling theory
for  classical fluids in \cite{EHvL-JSP-MC}, which is extended to
more general systems, and more general TCF's. 

What about the asymptotic decay of other single particle
properties, that are coupled to a conservation law? Consider the single site ($ \bn = \bzero$) correlation function
$\av{\bv_\bzero(t) \bv_\bzero(0)}$, initialized in a state of thermal equilibrium, which state may or may not be maintained by the dynamic evolution of the system.  The system considered has
{\it fixed} spins or velocity vectors $\bv_\bn$ at the sites $\bn =\{n_1,n_2,...,n_d\}$ of a $d-$dimensional cubic lattice with  nearest neighbor (n.n.) interactions conserving the total magnetization or momentum, $\sum_\bn \bv_\bn (t)$. For convenience we take periodic boundary conditions. Such models are  Glauber's model with Ising spins at zero temperature
\cite{Glauber}, or Lattice Gas Cellular Automata with $b$ 
velocities per site associated with the links \cite{LGCA}, or
granular fluid models with a discrete or continuous   velocity vector assigned to each site
\cite{Frenkel,BE98,PK-EBN-Springer}. In the following
 we consider as an example an exactly soluble  model of a granular fluid on a lattice whose LTT's are also covered by the results, to be derived in this letter. 

Here a pair of n.n. sites interacts at a rate $\kappa_0 $ such that
$\bv_\bn$ and $\bv_\bm$ are each replaced by their arithmetic mean.
Then the equation of motion for the mean value, $\av{\bv_\bn}
\equiv \bV_\bn$, can be written in appropriate units of length and
time as a discrete diffusion equation, i.e.
\be \Label{1a} 
d\bV_\bn /d\tau= \half \sum_{\bf a} [ \bV_{\bn +{\bf a}} -
\bV_\bn] = \half \Delta \bV_\bn
\ee
Here  $\tau = 2Dt/a^2$ is the rescaled time, $D=\kappa_0 a^2$ the diffusion coefficient, $a$ the lattice distance, ${\bf a}$ runs over n.n. sites, and $\Delta $ is the Laplace operator on a
$d-$dimensional cubic lattice. The exact solution
\cite{Glauber,PK-EBN-Springer} of this equation is $ \bV_\bn (\tau)=
\sum_\bm \bV_{\bn -\bm}(0) \: \prod_{i =1}^d [I_{m_{i}} (\tau)
e^{-\tau}]$, where $I_m (\tau)$ is the modified Bessel function.
The space-time correlation function $\av{v_{\bn x}(t) v_{\bzero
x}(0)}$ satisfies the same equation as $\av{v_{\bn x}}$, where $\av{\cdots}$ is an ensemble average over an arbitrary initial state. In the sequel all TCF's refer to initial states in thermal equilibrium. They are not necessarily spatially uniform. The single site TCF, $C_0(t) = \av{v_{\bzero x}(t)v_{\bzero x}(0)} /\av{v^2_{\bzero x}}$, follows by setting $\bn= {\bzero}$,  and yields the {\it exact} solution,
\be \Label{vacf}
C_0(t)  = \left( e^{-\tau} I_0(\tau) \right)^d \sim [ 2\pi
\tau]^{-d/2}
\ee
where the asymptotic equality gives its LTT. Here $C_0(t)$ is essentially the return probability of initial momentum to its points of origin. The same exact results apply  to the spin TCF in the Glauber model at zero temperature, and to the energy auto-correlation function $C_e(t) = \av{\delta \epsilon_\bn (t)    \delta \epsilon_\bn(o)}/\av{(\delta \epsilon)^2}$  in case the total energy $\sum_i \epsilon_i(t)$ is conserved, where $\delta \epsilon_\bn = \epsilon_\bn -\av{\epsilon}$. This function is again a return probability.  
 
In all previous models particles have only n.n. interactions, and are placed on lattice sites. These restrictions will be removed. When the particles are {\it moving}, as in fluids, one may also consider the tagged particle fluctuations, $j_2 = v^2 -\av{v^2}$, or more generally $j_{2k} =v^{2k} - \av{v^{2k}}$, and
$j_{2k+1} =v^{2k}v_x, ...$ with $k=0,1,...$, and even $j_0= \delta \epsilon$ in case the tagged particle has an additional internal degree of freedom with a discrete or continuous energy $\epsilon$.  Here the total energy of the system is conserved, and  we are interested in the LTT of these TCF's.  Such systems are classical fluids \cite{LTT-sim},
DPD fluids (Dissipative Particle Dynamics), which are mesoscopic  models \cite{DPD-e,DPD+e} with particle, momentum and possibly energy conservation, or the LBE (Lattice Boltzmann equation) method \cite{LBE}, which lacks energy conservation. The DPD fluid and LBE method can be considered as pre-averaged versions of respectively the Liouville equation and  the dynamic equations of lattice gas cellular automata \cite{LGCA}, where the rapid short-range fluctuations have been averaged out. 

In the DPD fluids the evolution equations for $\{\epsilon_i (t), \bv_i(t) \}$ are formulated as coupled Langevin equations with dissipative  and stochastic pair-interaction terms, having a finite interaction range $r_c$, and satisfying the fluctuation dissipation theorem. The stochastic interactions contain in general multiplicative white noise \cite{PRE-percol}. 
One may also quench the translational degrees of freedom, and freeze the particles at {\it fixed random} positions, or in {\it
fixed periodic} lattice configurations. In the former case a final average over quenched configurations has to be performed.  Here we consider only the quenched  DPD fluid with internal energy states,  referred to as DPD solid \cite{PRE-percol,LTT-RE}, where
dissipative and random interactions conserve the total internal energy. This quenched model of "interacting heat particles" is in several respects the dual model of the overlapping Lorentz gas\cite{PRE-percol}.

The purpose of this paper is to study the LTT's  of single- and multi-particle TFC's, $\av{J_b(t) J_b(0)}$,  in different  models,  and away from percolation and other critical points, using mode coupling theory. We start with fluid models. Here the LTT's can be concisely summarized  in  the Kadanoff-Swift formula \cite{EHvL-JSP-MC},
 \be \Label{MC}
\av{J_b(t) J_b(0)} \simeq \half \int_\bk \sum_{\lambda \mu} [A^{\lambda \mu}_b]^2 e^{(z^\lambda_k + z^\mu_k)t}
\ee 
where $\int_\bk =V^{-1}\sum_\bk $. In the thermodynamic limit
$\int_\bk$ is replaced by $(2 \pi)^{-d} \int d\bk$.  The $\lambda\mu$-sum extends over the complete set of hydrodynamic modes $\{a^\lambda_\bk \}$ of the system. The  amplitude $A_b^{\lambda\mu} = (J_b |a^\lambda_\bk a^\mu_{-\bk}) \equiv V^{-1}\av{J_b a^\lambda_\bk a^\mu_{-\bk}}$ represents the component of $J_b$ parallel to a product of two modes. These modes are linear combinations of Fourier transforms of conserved densities, and are normalized as $( a^\lambda_\bk| a^\mu_{\bk}) =\delta_{\lambda\mu}$. The scalar products are defined as $(A_\bk|B_\bk) = V^{-1} \av{A_\bk B_{-\bk}}$, averaged over an equilibrium ensemble. All TCF's considered here approach zero for large times, which implies that $(J_b|a^\lambda_\bk) =0$.
 
In a {\it classical fluid} there are $(d-1)$ diffusive shear    modes $(\lambda =\eta_i: i=1,2, \cdots,d-1)$, being the components
of  the tranversal flow field ${\bf u}_{\bk \bot} = {\bf u}_\bk - \hat{\bf k} \hat{\bf
k}\cdot {\bf u}_\bk $, with decay rates $z^\eta_k = -\nu k^2$ and shear viscosity $\eta= \rho \nu$, and
one diffusive heat mode $a^H_k$ with decay rate $z_\bk^H = - D_T
k^2$ and heat diffusivity $D_T$. In addition there are two damped
propagating sound modes $(\lambda =\sigma = \pm)$ with decay rate
$z^\sigma_k = -i \sigma c_0 k -\half \Gamma_s k^2$, where $c_0$ is
the speed of sound and $\Gamma_s$ the  sound damping constant. For
the explicit expressions of fluid modes  $a^\lambda_\bk$ in terms
of conserved densities, and the calculation of the dominant LTT's we refer to \cite{EHvL-JSP-MC}.  
 
A special case are the {\it single} particle TCF's $\av{j_b(t) j_b(0)}$ in fluids. Here the mode coupling formula necessarily involves the Fourier mode of the {\it tagged } particle density, i.e. the self diffusion mode, $a^s_\bk = n^s_\bk = \exp[-i\bk\cdot\br_1]$, where $i=1$ is the label of the tagged particle. Consequently the following replacements have to be made in \Ref{MC}: $ \mu \to s $ and $\half \sum_{\lambda \mu} \to \sum_\lambda $. Moreover, if one of the labels equals $s$, then the amplitude $A_b^{\lambda s} = (j_b|a_\bk^\lambda a_{-\bk}^s)  \equiv  \av{j_b a_\bk^\lambda a_{-\bk}^s}$ without a factor $1/V$. To extract the long time behavior from \Ref{MC} we change the
integration variable $\bk$ to ${\bf q} /\sqrt{t}$, and take the
long time limit, using $a^\lambda_\bk \to a^\lambda_\bzero$ and $n^s_\bk \to n^s_\bzero=1$. So, the amplitude of the single particle function simplifies to $ A_b^{\lambda s} = \av{j_b |a^\lambda_0}$, and depends only on the direction $\hat{\bk}$ of the wave vector $\bk$. The final result for  the LTT becomes,
\be \Label{3}
\av{j_b(t) j_b(0)} \simeq \sum_\lambda \overline{[\av{j_b|a^\lambda_0} ]^2}/ [4 \pi (D_\lambda +D)t]^{d/2}
\ee
where only diffusive modes ($\lambda =(\eta_i,H)$) contribute.
The overline indicates an average over the solid angle
$\hat{\bk}$. As the calculations of $A_b^{\lambda
s}$ are similar to those in [2b], we simply
quote the final results for the LTT in the equilibrium TCF's of classical fluids, $C_b(t)= \av{j_b(t)j_b(0)} /\av{j_b^2}$ with $b = \{1,2,3,4,...\}$, and $j_b = (v_{x},v^2-\av{v^2}, v^2 v_x, v^4-\av{v^4})$, i.e.
\ba \Label{4}
C_1(t)  &\simeq& [(d-1)/nd]  [4 \pi (\nu +D)
t]^{-d/2}
\nn C_2(t) &\simeq&  ({\cal C}^0_v/n {\cal C}_p)[4 \pi (D_T +D) t]^{-d/2}
\ea
with ${\cal C}_p$ the specific heat at constant pressure,
and ${\cal C}^0_v =\half d k_B$. One similarly shows that odd-in-$v$, c.q. even-in-$v$ correlations are at {\it large times}
proportional to $C_1(t)$  and $C_2(t)$, i.e.
$C_3(t)  \simeq  [(d+2)/(d+4)]C_1(t)$, $C_4(t)  \simeq [(d+2)/(d+3)]C_2(t)$, etc for $k=5,6,\cdots$.
The result for the VACF in hard sphere or Lennard-Jones fluids is well known \cite{LTT-sim,EHvL-JSP-MC}. The remaining ones are new. The above results also disprove the misconception that momentum conservation is necessary for the existence of LTT's in classical fluids. The results in \Ref{3}-\Ref{4} for classical fluids also apply to  DPD fluids with energy  and momentum conservation\cite{DPD+e}. In DPD fluids without energy conservation \cite{DPD-e}  only the relations for odd-$k$ values apply because the heat mode is absent, and the system is  thermostatted instantaneously.

Next we want to extend these results to the random DPD solid.
The mode coupling results \Ref{MC} and \Ref{3} for TCF's in fluids do in general not apply to systems with quenched disorder \cite{Machta}. There they apply only to TCF's $C(t|X)$, calculated in a non-uniform equilibrium state, corresponding to a single quenched configuration $X$, where $A^{\lambda\mu}(X)$ and $D_T(X)$ in \Ref{MC} and \Ref{3} depend on the configuration $X$. To obtain the full correlation, $C(t) =\av { C(t|X)}$, a subsequent average over all quenched configurations has to be performed. In general the $X-$dependence of $A^{\lambda\mu}(X)$ and $D_T(X)$ is not known explicitly. A more phenomenological derivation of a mode coupling formula for diffusive systems with quenched disorder has been presented in Ref.\cite{Machta}, and one may try to extend that method to the heat conducting random DPD solid. However, we will not follow that route here, but investigate only those special cases for which the LTT of $C(t|X)$ can be determined explicitly from \Ref{MC}.

Consider first the {\it single} particle energy correlation $C_e(t|X) = \av{\delta\epsilon_i (t) \delta\epsilon_i(0)}_X$ in the DPD solid, where the heat mode, $a^H_\bk (t) = a^H_\bk (0) \exp[- tk^2D_T(X)]$, is the only slow macroscopic mode, and $a^H_\bk = e_\bk /\sqrt{(e_\bk|e_\bk)}$. In the long time limit (where $\bk \to 0$) the relevant amplitude $A^{Hs}(X) = \av{\delta\epsilon_1 a^H_\bk a^s_{-\bk}}_X \simeq \av{\delta\epsilon_1 a^H_\bzero}_X = \sqrt{V \av{(\delta \epsilon)^2}/N}$ where $a^s_\bk =
\exp[ - i \bk \cdot \br_1]$ is a frozen mode with $z^s_k =0$. Inserting these results in \Ref{3}, and performing the $X-$average yields the LTT,
\be \Label{X}
C_e(t) \simeq \frac{V}{N}\left(\frac{1}{4 \pi D_T t}\right)^{d/2} =\frac{1}{\rho[4 \pi t^*]^{d/2}}
\ee
Here $\av{D_T(X)} =D_T$, and fluctuation corrections of relative order $\av{(\delta D_T(X))^2}/D_T $ have been neglected. Moreover $t^*=Dt/r_c^2$ and $\rho =N r^2_c/V$ is proportional to the mean number of particles inside an interaction sphere of radius $r_c$. If the particles are put on a lattice (no disorder), the LTT in $C_e(t)$  applies as well to the lattice version of the DPD solid. Both on-lattice and off-lattice computer simulations of the LTT in $C_e$  are in {\it excellent} agreement \cite{LTT-RE} with the theoretical prediction \Ref{X}.

Next we study the $N-$particle correlation $C_Q(t)=\av{ Q_x(t) Q_x(0)}/V$. The random DPD solid sustains a microscopic heat flux, $Q_x =Q_D +Q_R$, with a dissipative (D) and a random part (R), as given explicitly in \cite{PRE-percol}. We start with 
$ Q_D(t)= \kappa_0 \sum_{i < j} w(r_{ij}) {r}_{ij,x}
({\epsilon_j}(t) - {\epsilon_i}(t))$. This sum of dissipative pair interactions represents the instantaneous exchange of energy over a distance $r_{ij}$ through the interactions (collisional transfer). The model parameter $\kappa_0$ represents the interaction frequency, and the range function is a step function, vanishing for $r > r_c$, and normalized such that $\int d\br w(r) =r_c^d$. Consequently $\kappa_0 (\epsilon_j-\epsilon_i) w(r_{ij})$ is the rate of energy transfer between the interacting pair $(ij)$ with $r_{ij} <r_c$. The corresponding TCF is $C_D(t) = V^{-1} \av{ Q_D(t) Q_D(0)}$. Next we consider the stochastic part of the heat flux $Q_R$, and the corresponding correlation $C_R(t)$ $ = V^{-1} \av{Q_R(t) Q_R(0)}$ and cross-correlations $C_{DR}(t)$ and $C_{RD}(t)$. For our present purpose it is sufficient to note  that $Q_R$ is  linear in the Langevin force, $\tilde{F}_{ij}(t)=  - \tilde{F}_{ji}(t)$, with coefficients depending on $\epsilon_i$  and $\epsilon_j$ (multiplicative noise), where $\av{\tilde{F}_{ij}(t)}=0$ and $\av{\tilde{F}_{ij}(t) \tilde{F}_{kl}(t^\prime)}$ $= \delta (t-t^\prime)$ $ (\delta_{ik}\delta_{jl} -\delta_{il}\delta_{jk})$. The latter property guarantees that $C_R(t)$ is delta-correlated in time, $C_R(t) \simeq (2 \lambda_\infty /k_B \beta^2) \delta (t)$. Here $k_B$ is Boltzmann's constant, $\beta = 1/k_bT$ the inverse temperature, and $\lambda_\infty$ the heat conductivity in mean field approximation ($\rho \to \infty$), as calculated in \cite{PRE-percol}. The cross-correlations, $C_{DR}$ and $C_{RD}$, vanish being linear in $\tilde{F}_{ij}$.

The heat conducting DPD solid is in fact a diffusive system with static disorder (see Ref.[12]) with heat conductivity $\lambda = k_B\beta^2 \int_0^\infty dt C_Q(t)$. We apply the mode coupling theory developed for such systems in Ref.[12] to calculate the LTT of $C_Q(t)$. To do so the local concentration, $c(\br,t)$ in the fluctuating diffusion equation of Ref.[12] needs to be replaced by the energy density $e(\br,t)$, to obtain
\be 
\partial_te(\br,t)= {\bnabla} \cdot {\boldmath{\lambda}}(\br,X) \cdot {\bnabla} [e(\br,t)/\psi(\br,X) ],
\ee 
where $\psi(\br,X) = {\cal C} n(\br)$ and ${\cal C} $ is the specific heat per DPD particle. The tensor $\lambda^{\alpha\beta}(\br,X)$ is the spatially fluctuating heat conductivity tensor with $ \av{\lambda^{\alpha\beta}(\br,X)} = \lambda \delta_{\alpha\beta}$, and the heat diffusivity is $D_T = \lambda /\av{\psi} =\lambda /{\cal C} n $. Following the derivation of Ref. [12] one finds for the LTT,
\be
C_Q(t) \simeq C_D(t) \simeq - \pi n k_B T^2 {\cal C} \Delta /(4 \pi D_T t)^{1+d/2},
\ee
where $\Delta$ is the mean square fluctuation in the $(\bk =0)$ Fourier components of $\delta D^{\alpha\beta}(\br,X) = \delta \lambda^{\alpha\beta}(\br,X)/[{\cal C} n]$, i.e.
\be
\Delta = (1/dV) \av{\delta D_0^{\alpha\beta}(\br,X)\delta D_0^{\beta \alpha}(\br,X)}
\ee 
with implied summation convention for repeated indices. The LTT $\sim t^{-1-d/2}$ of $C_D$ in the random solid has the same structure as the VACF in the Lorentz gas, and both LTT's vanish when the fixed particles are filling the sites of a periodic lattice [13], because $\Delta=0$.

In the present system one has for {\it large densities} 
the explicit expression for the fluctuating heat diffusivity $D^{\alpha\beta}(\br,X) = (\lambda / nV)\sum_{i<j} w(r_{ij}) r_{ij,\alpha} r_{ij,\beta}$, and the quantity $\Delta$ can be calculated, yielding the LTT at large densities, 
\be 
 \frac{C_D(t)}{C_D(0)}  \simeq  - \frac{ \pi}{\rho}\left( 
 \frac{d+2}{d+4}\right) \left( \frac{1}{ 4 \pi t^*} \right)^{1+d/2}.
\ee

 Do the above results have any implications for {\it dense} classical fluids, say in the vicinity of the triple point? We propose the following scenario. In such dense systems the motion of the particles is quite restricted - somewhat comparable to a quenched system -- and collisional transfer of momentum and energy is the dominant transport mechanism. Consider for example the microscopic stress tensor. If the rate of energy transfer between the particle pair $(ij)$ in $ Q_D$ is replaced by the rate of momentum transfer, i.e. the inter-particle force,
$F_{ij,y} = - \partial V(r_{ij})/\partial r_{ij,y}$, then $Q_x$ becomes the collisional transfer component $S^c_{xy}$ of the stress tensor. This suggests that in dense fluids  the stress correlation function has a LTT, $C^c_S(t) = V^{-1} \av{S^c_{xy}(t) S^c_{xy}(0)} \sim A_c t^{-1-d/2}$, similar to the velocity auto-correlation function in the Lorentz gas \cite{Machta}. Standard mode coupling theory for fluids only predicts that the TCF $C^k_S(t)$ of the kinetic stresses, $S^k_{xy} =\sum_i mv_{ix} v_{iy}$, has a LTT $\sim A_k t^{-d/2}$, \cite{EHvL-JSP-MC,ladd,Frenkel}, and  that no such tail is present in $C^c_S(t)$.

Standard MD simulations of TCF's in hard sphere fluids \cite{ladd} have not been able to establish a LTT $\sim t^{-d/2}$ in $C^c_S$, as is consistent with theory. However,
the non-equilibrium MD simulations \cite{evans} for small Lennard-Jones systems under constant shear rate $\dot{\gamma}$ seem to suggest an (intermediate ?) LTT of $C^c_S(t) \sim A_c t^{-d/2}$ with $A_c$ one or two order of magnitude larger than the value $A_k$, predicted by the theory for $C_S^k$. Here determination of the LTT involves two non-uniform limits $(\dot{\gamma} \to 0, t \to \infty)$. A large intermediate tail $\sim t^{-d/2}$ may be a crossover phenomenon  for $t < t_{cross}(\dot{\gamma})$ at a small, but non-vanishing $\dot{\gamma}$, where the collisional transfer stress correlation in dense fluids is showing the $t^{-d/2}-$tail of Burnett-type correlations, similar to those derived for Lorentz gases \cite{Machta}.

It would be of great interest to test the proposed scenario for LTT's in TCF's in dense classical fluids by performing detailed 
computer simulations of $C_Q$ in DPD solids and fluids, as well as by developing a quantitative analysis of $C_Q$ for general densities. Promising alternatives of measuring these LTT's, particularly in 1-D, are offered by simulating the Helfand moments \cite{helfand}.
 
\hspace{-10mm} 
\section*{Acknowledgements}
\hspace{-3mm}
The author acknowledges stimulating discussions and/or correspondence with E. Ben-Naim, J. Machta, J.R. Dorfman, H. van Beijeren, M. Ripoll and I. Pagonabarraga.

\end{document}